\documentclass[twocolumn]{aastex62}
\usepackage{lineno}
\usepackage{hyperref}
\shorttitle{On-the-Fly Pulsar Mapping} 
\shortauthors{Swiggum \& Gentile}

\submitjournal{the Astronomical Journal}
\begin{document}

\title{On-the-Fly Mapping of New Pulsars}

\author[0000-0002-1075-3837]{J.~K.~Swiggum}
\affiliation{Center for Gravitation, Cosmology and Astrophysics, Department of Physics, University of Wisconsin-Milwaukee, P.O. Box 413, Milwaukee, WI 53201, USA}

\author{P.~A.~Gentile}
\affiliation{Department of Physics and Astronomy, West Virginia University, P.O. Box 6315, Morgantown, WV 26505}
\affiliation{Center for Gravitational Waves and Cosmology, West Virginia University, Chestnut Ridge Research Building, Morgantown, WV 26505}

\correspondingauthor{J.~K.~Swiggum}
\email{swiggumj@uwm.edu}

\begin{abstract}
Current single dish, low-frequency radio pulsar surveys provide efficient sky coverage, but poor localization of new discoveries. Here, we describe a practical technique for rapidly localizing pulsars discovered in these surveys with on-the-fly mapping and provide code to facilitate and formalize its implementation. As a proof of concept, we alter the positions of four test sources and use the Green Bank Telescope (GBT) 350\,MHz receiver to recover source positions within $\approx1-3\,\arcmin$ of their true values, compared to an $18\arcmin$ error radius for new discoveries. Achieving similar precision with a traditional gridding strategy using the GBT requires $2-3$ times as much telescope time (including overhead), multiple receivers and relies on assumptions about the pulsars' spectral indices. For one of our test sources (PSR J1400$-$1431), this method revealed a discrepancy with the initial, published position, prompting additional follow-up and an improved timing solution. Rapid localization is important for improving data quality and providing flexibility in choice of center frequency for future timing observations -- both of which facilitate evaluating new millisecond pulsars for potential inclusion in pulsar timing arrays.
\end{abstract}

\keywords{methods: data analysis --- pulsars: general}

\section{Introduction}

The choice of center frequency for single dish radio pulsar surveys is critical.  Low frequencies maximize survey speed in covering large areas of the sky and help take advantage of pulsars’ steep spectral energy distributions, while high frequencies minimize the adverse effects of sky temperature, scattering, and dispersive smearing. Since a telescope's angular beam size is inversely proportional to observing frequency ($f$) and its diameter ($D$), the time required to conduct all-sky surveys with the most sensitive (largest diameter) telescopes at frequencies $>1$\,GHz becomes prohibitive. For this reason, recent all-sky pulsar surveys with the Green Bank Telescope (GBT) and Arecibo Observatory (AO) -- two of the largest radio telescopes in the world, both with $D>100$\,m -- have been carried out at center frequencies, $f\approx350$\,MHz \citep{slr+14,dsm+13}.

Although low-frequency surveys are more efficient in survey speed, discoveries are poorly localized. For example, the Green Bank North Celestial Cap \citep[GBNCC;][]{slr+14} survey and the 350\,MHz Drift Scan Survey before it \citep{blr+13,lbr+13,rsm+13} used the GBT's 350\,MHz receiver, which is sensitive to an angular region on the sky $36\arcmin$ wide. With an initial error circle that large, conducting follow-up observations at higher frequencies without further refinement is unreliable given that the GBT beam sizes at 820\,MHz and 1.5\,GHz are $15\arcmin$ and $9\arcmin$ wide respectively. This can delay multi-frequency follow-up, which is important for high-precision timing -- particularly for millisecond pulsars (MSPs) in pulsar timing array (PTA) experiments \citep{abb+15,dcl+16,rhc+16,vlh+16}. These experiments require MSPs to be monitored at multiple frequencies in order to precisely measure and correct for dispersive delays and other interstellar medium effects, which change over time \citep{jml+17}.

Multi-frequency timing precision must be assessed before adding new MSPs to PTAs. In order to do this properly with the GBT, an MSP position must be known to $\lesssim3\arcmin$ (i.e. within a 1.5\,GHz beam radius from the true position). Pulsar timing provides sub-arcsecond position measurements, but it requires $\simeq1$\,year of monthly observing and a high-cadence session to achieve phase connection and yield results. Conducting multi-frequency follow-up on shorter timescales requires other methods.

\subsection{Gridding}
A commonly-used approach to incrementally improve pulsar localization is called {\it gridding} and involves tiling the discovery error region with multiple scans to refine a pulsar's position in stages \citep{mhl+02}. Oftentimes gridding observations are carried out at higher observing frequencies (smaller wavelengths, $\lambda$) to more significantly improve localization \citep[e.g.][]{lbr+13} since beam size is proportional to $\lambda$ (see Equation \ref{eq:hpbw}).

Considering a pulsar discovered with the GBT at 350\,MHz, for example, gridding typically involves tiling the $36\arcmin$-wide beam with seven 820\,MHz pointings. Then, one or more detections at the higher frequency are used to determine a central position for a second round of seven grid pointings carried out at 1.5\,GHz to further refine the position. At each stage, if multiple grid points produce detections, an average position is calculated, weighted by detection signal-to-noise ratios. In the best case scenario, progressive stages of this process improve the pulsar's error radius to $\lesssim5\arcmin$, then $\lesssim2\arcmin$ respectively.

This can be logistically complicated due to the overhead time required for receiver switches (10$-$15\,minutes). Also, gaining access to these receivers may require dedicated proposals. Furthermore, pulsars are steep-spectrum objects \citep{blv+13}; increasing the observing frequency by a factor of $\approx2$ decreases the expected pulsar flux by a factor of $\approx3$. Even after considering the additional bandwidth and improved system temperatures at higher observing frequencies, pulsars' steep spectral indices require increased observing time for individual grid points to ensure detections. For a two-minute discovery scan at 350\,MHz, grid point scans at 820\,MHz and 1.5\,GHz are typically $\approx5$\,minutes to be safe. Including overhead time spent switching receivers and slewing, the entire process can take $1.5-2$\,hours per source, resulting in source localization of $\lesssim2\arcmin$.

In this paper, we outline \--- in some cases \--- a more effective approach for rapid localization of new pulsars using on-the-fly (OTF) mapping. OTF mapping is not a novel technique (commonly used to measure positions for continuum sources with single dish telescopes), but one that is under-utilized in the pulsar community and is especially convenient for localizing pulsars discovered in drift scan surveys and with telescopes that do not have multi-beam capabilities. This study is a proof of concept, motivated by the need for rapid assessment of new MSPs for inclusion in PTAs, and meant to facilitate the process.

In \S\ref{sec:method}, we describe the method, observations of four test sources, and our procedure for measuring position offsets; \S\ref{sec:results} describes the precision of recovered positions resulting from our test observations and a notable result for millisecond pulsar J1400$-$1431 -- its measured coordinates were discrepant with the previously published values, prompting additional follow-up and an improved timing solution. Finally, we discuss systematic errors that should be considered and conclude in \S\ref{sec:conclusion}.

\section{On-the-Fly Mapping}\label{sec:method}
\subsection{Motivation}
Shortly after discovering a new pulsar in a single-dish survey, the position is only known to within the telescope's half-power beam width ($\theta_{\rm HPBW}$). A detailed description in Essential Radio Astronomy \citep[][hereafter ERA]{cr+16} shows that $\theta_{\rm HPBW}\propto\lambda/D$, where $\lambda$ is the observing wavelength and $D$, the telescope diameter. The simplest case for calculating this constant of proportionality involves assuming that the telescope aperture ($D$) is uniformly illuminated. However, most feeds do not illuminate an aperture uniformly, so using a more realistic cosine-tapered illumination pattern,
\begin{equation}
\theta_{\rm HPBW}\approx 1.2\,\frac{\lambda}{D}
\label{eq:hpbw}
\end{equation}
(see ERA 3.96 for the full derivation). Although the resulting power pattern has a rather complicated functional form, it is well-approximated by a Gaussian  with full-width half maximum equal to $\theta_{\rm HPBW}$, i.e.
\begin{equation}
P(\theta)\propto\exp\bigg[-4\ln2\,\bigg(\frac{\theta}{\theta_{\rm HPBW}}\bigg)^2\bigg].
\label{eq:gausspower}
\end{equation}
Here the beam's power pattern ($P$) is expressed as a function of $\theta$, an angle measured from the beam center. On-the-fly (OTF) mapping at a constant rate across an unresolved point source will produce an intensity pattern, or {\it beam profile}, with the same functional form. The beam profile's maximum amplitude is then determined by the source's flux density, the telescope's sensitivity, and the mapping impact parameter (closest approach angle between beam center and source position). For mapping traces with small -- but non-zero -- impact parameters ($\ll$~1 rad), the maximum amplitude of the resulting beam profile will be lower than a situation where the telescope's boresight maps directly across the source's position. However, regardless of the size of the impact parameter, $\theta_{\rm HPBW}$ will not change, so the beam's angular width is a fixed parameter in a Gaussian beam model, which is fit to the measured beam profile to determine a source's actual position and its uncertainty.

\subsection{Observations}\label{sec:obs}
On September 2, 2014 we used the 350\,MHz receiver and the Green Bank Telescope (GBT) to observe four test sources, including B1919+21 \--- a relatively bright, slow pulsar in the northern hemisphere \--- and three others discovered by high school students participating in the Pulsar Search Collaboratory \citep[PSC;][]{rhm+10}, PSRs J1400$-$1431, J1822+0155, and J1930$-$1852. Observations were conducted in incoherent search mode using the Green Bank Ultimate Pulsar Processing Instrument \citep[GUPPI;][]{drd+08}, recording 4096 channels every 81.92~$\mu$s. 

The chosen test sources had presumably already been localized to sub-arcsecond precision in earlier pulsar timing campaigns \citep{rsm+13,srm+15}. To test how well OTF mapping could recover positions measured with pulsar timing ($\alpha$, $\delta$), we changed them by known amounts ($\Delta\alpha$ and $\Delta\delta$; see Table \ref{tab:offsets}) and conducted 15\,min scans across $2^{\circ}$ ($\approx4\times\theta_{\rm HPBW}$) in right ascension and declination directions, using {\tt RALongMap} and {\tt DecLatMap} functions respectively (see GBT observing manual\footnote{\url{https://science.nrao.edu/facilities/gbt/observing/GBTog.pdf}}). These scans were centered on altered coordinates $\alpha+\Delta\alpha$ and $\delta+\Delta\delta$ and a $\cos(\delta)$ factor was used to calculate angular offsets in the $\alpha$-direction due to the fact that lines of constant right ascension on the celestial sphere converge for $|\delta|>0$.

\subsection{Data Reduction}\label{sec:reduction}
Scans in $\alpha$ and $\delta$ directions were processed identically, using standard routines from {\tt PRESTO}\footnote{\url{http://www.cv.nrao.edu/~sransom/presto/}} \citep{rem+02} and code described below to measure a source's offset from the center catalog position in each direction can be found at \href{http://doi.org/10.5281/zenodo.1346351}{10.5281/zenodo.1346351}.

\begin{figure*}
    \centering
    \gridline{\fig{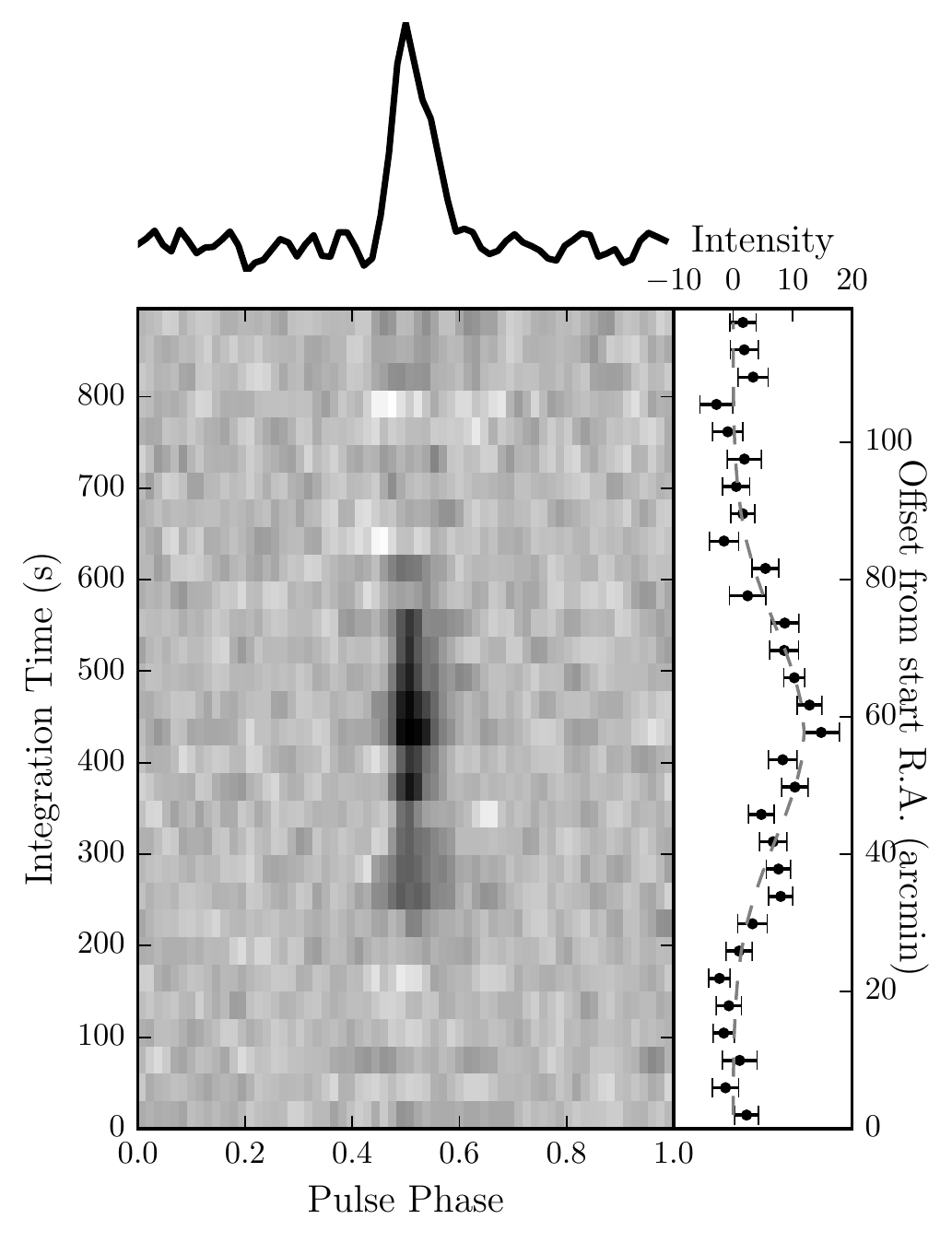}{0.5\textwidth}{(a) PSR J1400$-$1431 (Right Ascension)}
    		  \fig{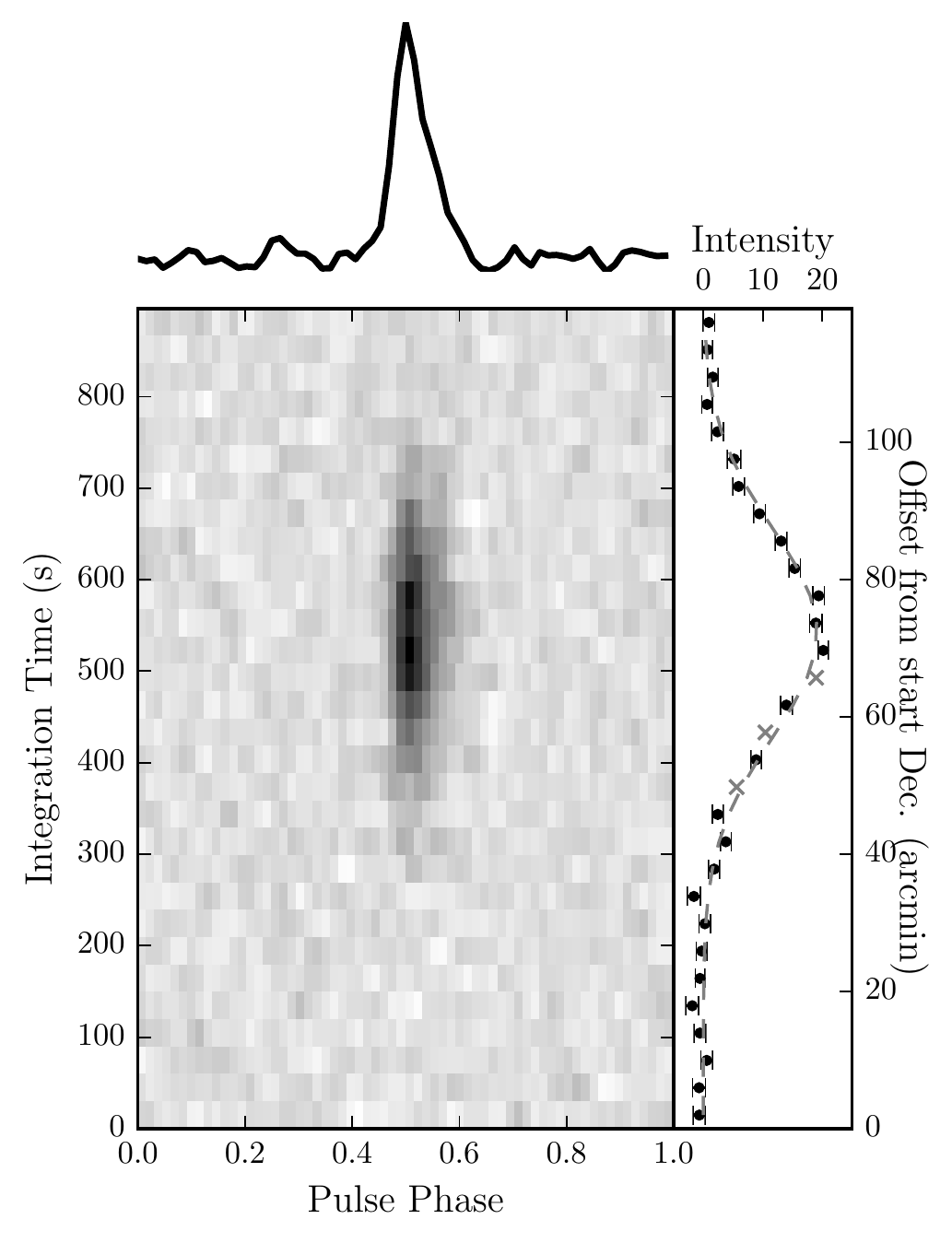}{0.5\textwidth}{(b) PSR J1400$-$1431 (Declination)}}
    \caption{Right ascension (a) and declination (b) OTF mapping scans for PSR J1400$-$1431. For each scan, the left subplot shows pulse phase versus time; the beam profile (black points with uncertainties) in the right subplot shows how the measured intensity (arbitrary units) traces the telescope beam power pattern as the beam slews across the source. The resulting folded pulse profile is shown on top of each panel. The gray dashed line represents the best-fit beam model whose mean value corresponds to the measured position offset. Gray $\times$s indicate beam profile measurements rejected due to their uncertainties lying outside a threshold determined by the interquartile range.\label{fig:fold}} 
\end{figure*}

Radio frequency interference (RFI) was zapped automatically with {\tt rfifind} and data files were dedispersed, then folded with {\tt prepfold}, using 64 phase bins and 30 sub-integrations (see Figure \ref{fig:fold}). With {\tt pygaussfit.py}, the resulting folded profiles were modeled with one or more Gaussian components to create noiseless templates.

Each template was normalized to have a maximum value of one. On/off-pulse profile bins were determined by choosing an appropriate threshold near zero, then identifying bins from the noiseless template whose intensities were respectively above/below the chosen threshold. Before marginalizing over frequency, the bandpass shape was removed by subtracting the median value of each frequency channel for all profiles. Profiles were further cleaned by fitting and subtracting low-order polynomials ($2<N<6$) from their off-pulse baselines. Finally, profiles were scaled by a global standard deviation of the off-pulse bins.

Amplitudes for each pulse profile were calculated simply by scaling the noiseless template described earlier and their uncertainties were estimated with $\sigma_{\rm off}/\sqrt{N_{\rm on}}$, where $\sigma_{\rm off}$ is the standard deviation of an individual profile's off-pulse bins and $N_{\rm on}$ is the number of on-pulse bins. We verified that uncertainties in amplitude fits scaled identically for simulated profiles with varying amplitudes, amounts of noise, duty cycles, and pulse widths. To mitigate the effect of amplitude measurements with unusually small uncertainties (sometimes this occurs as a result of masking RFI), $\log_{10}$ amplitude uncertainties below the median by more than twice the interquartile range were heavily down-weighted so that they did not affect beam profile fitting. Offsets \--- calculated for each profile using corresponding timestamps and the telescope slew rate \--- were measured relative to the start position, taken directly from the file header.

Gaussian-approximated 350\,MHz beam models of known width (Equation \ref{eq:gausspower}; $\theta_{\rm HPBW}=36$\,\arcmin) were fit to resulting beam profiles to accurately measure offsets, $\Delta\alpha_{\rm meas}$ and $\Delta\delta_{\rm meas}$ respectively (see Table \ref{tab:offsets}). Fitting was carried out in $\alpha$/$\delta$ directions independently using {\tt optimize.curve\_fit}, a python implementation of a non-linear least squares fitter in the {\tt scipy} library, and the beam profile mean and amplitude were included as fit parameters. An initial fit was used to calculate a reduced chi-squared ($\chi^2_{\rm red}$) value, then beam profile amplitude errors were multiplied by ${\rm EFAC}=\sqrt{\chi^2_{\rm red}}$ to achieve $\chi^2_{\rm red}=1$. Fit uncertainties were initially estimated using diagonal elements from the covariance matrix that was returned as part of the fitting routine. Next, a bootstrapping technique \citep{efron+1979} was employed, drawing random sets of beam profile amplitudes with replacement for 10,000 trials. The mean offsets measured using the bootstrapping technique were nearly identical to those measured by fitting beam profiles directly. Standard deviations of the bootstrapping results were also similar to beam profile fitting uncertainties calculated using the covariance matrix, but in some cases, the former were as much as 25\% larger than the latter. Bootstrapping results were adopted since their slightly more conservative uncertainties put recovered offsets in better agreement with true source positions. 

In addition to the one-dimensional (1D) fits in $\alpha$ and $\delta$ described here, we also employed a 2D fit using amplitudes measured from both scans simultaneously and a 2D Gaussian beam model with $\theta_{\rm HPBW}=36\arcmin$. Again, a bootstrapping technique was used to measure offsets and uncertainties and the bootstrapping results from both methods are shown in Table \ref{tab:offsets}. A full comparison of recovered positions (both with and without bootstrapping) is illustrated in Figure \ref{fig:recovered}.

\begin{deluxetable*}{lcrrrrrrc}
 \tablewidth{0pt}
\tablecaption{Recovered Position Offsets}
\tablehead{
\colhead{Source} & \colhead{Slew Rate (\arcmin\,s$^{-1}$)} & \colhead{$\Delta\alpha$ (\arcmin)} & \colhead{$\Delta\delta$ (\arcmin)} & \colhead{$\Delta\alpha_{\rm rec,1D}$ (\arcmin)} & \colhead{$\Delta\delta_{\rm rec,1D}$ (\arcmin)} & \colhead{$\Delta\alpha_{\rm rec,2D}$ (\arcmin)} & \colhead{$\Delta\delta_{\rm rec,2D}$ (\arcmin)} & \colhead{$\xi$ (\arcmin)}} 
\startdata
J1400$-$1431\tablenotemark{a} & 0.133 & 2.9 & $-$5.0 & $-2.0\pm1.9$ & $12.3\pm0.5$ & $-1.5\pm1.6$ & $12.8\pm0.6$ & $-$ \\
J1400$-$1431\tablenotemark{b} & 0.133 & 0.0 & 0.0 & $-0.7\pm1.9$ & $0.9\pm0.5$ & $-0.2\pm1.6$ & $1.3\pm0.5$ & 0.4\\
J1822+0155 & 0.133 & $-$11.0 & 8.0 & $12.7\pm0.9$ & $-6.8\pm1.3$ & $11.5\pm0.9$ & $-8.1\pm1.4$ & 0.3 \\
J1930$-$1852 & 0.133 & 3.8 & $-$3.0 & $-4.4\pm0.8$ & $2.2\pm0.8$ & $-4.5\pm1.0$ & $2.6\pm1.5$ & 0.6 \\
B1919+21 & 0.133 & 10.2 & $-$10.0 & $-10.3\pm0.2$ & $7.5\pm0.3$ & $-12.2\pm0.6$ & $6.2\pm0.8$ & 0.2 \\
B1919+21 & 0.666 & 0.0 & 0.0 & $-3.5\pm0.8$ & $1.1\pm0.6$ & $-3.1\pm0.6$ & $1.0\pm0.5$ & 0.7
\enddata
\tablecomments{Altered source positions and recovered offsets with uncertainties using both 1D and 2D beam profile fits. The difference between the target source position entered into the telescope control scripts and the actual central position of the OTF map scans is given by $\xi$ for each source.}
\tablenotetext{a}{Recovered offsets measured relative to the initially incorrect position published in \cite{rsm+13} with injected offsets $\Delta\alpha$ and $\Delta\delta$: $14^{\rm h}00^{\rm m}$42\fs34 $-$14\degr43\arcmin14\farcs43.}
\tablenotetext{b}{Recovered offsets measured relative to the corrected timing position, published in \cite{skm+17}: $14^{\rm h}00^{\rm m}$37\fs00370 $-$14\degr31\arcmin47\farcs0422.}
\label{tab:offsets}
\end{deluxetable*}

\section{Results}\label{sec:results}

\subsection{Corrected Position for PSR J1400$-$1431}\label{sub:1400}
For several other test sources, we found that the recovered positions were within 1-$\sigma$ uncertainties from published timing positions. This was not the case for PSR J1400$-$1431, where for both 1D and 2D fits, $\Delta\delta_{\rm rec}>12'$, more than $7\arcmin$ larger than the injected offset, suggesting the true source position differed significantly from that published in \cite{rsm+13}, motivating further timing analysis for J1400$-$1431.

The OTF mapping position helped establish a fully phase-connected timing solution for J1400$-$1431  \citep{skm+17}, covering a longer timespan than that published by \cite{rsm+13}, based on $\sim$7~months worth of timing data. Comparing the new position to the earlier value, they differed by $6.7\arcmin$. This discrepancy comes partially from the fact that J1400$-$1431 was discovered at the edge of the GBT 350\,MHz beam (and subsequently, was only detected in one 820\,MHz grid pointing) and is partially due to covariance between position and spin-down parameters in the initial timing solution. The improved timing position is $21.5'$ from the discovery position and $11.1'$ from the first 820\,MHz grid position. This offset explains earlier difficulties detecting J1400$-$1431 over the first year of observations and particularly, during earlier tests of J1400$-$1431's suitability for inclusion in pulsar timing arrays (PTAs). Recent test observations using the improved timing position suggest that it still has too steep a spectrum to be reliably detected at $>1$\,GHz, and is therefore not a suitable addition to PTAs.

In Table \ref{tab:offsets}, J1400$-$1431 is listed twice to show $a$) the discrepant offsets measured relative to the initially incorrect position and $b$) the recovered position's close proximity to the corrected timing position ($\approx1\arcmin$).

\begin{figure}[t]
\includegraphics[width=0.5\textwidth]{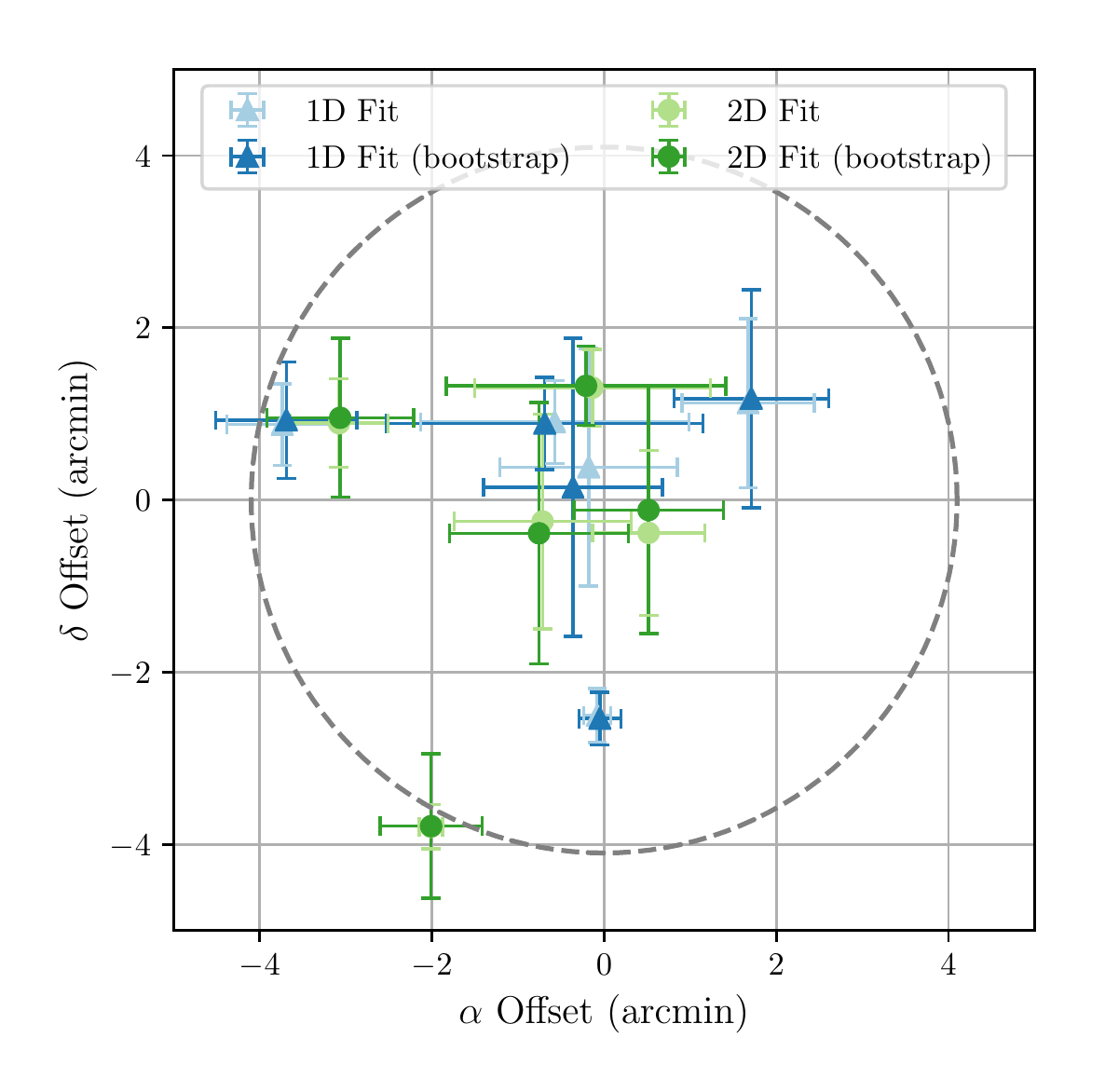}
\caption{Recovered $\alpha$ and $\delta$ offsets plotted relative to actual values (origin) for each test pulsar. Different colors/symbols differentiate between results from independent 1D fits (blue triangles) and simultaneous 2D fits (green circles). Light and dark points differentiate between direct fits to beam profiles and results from the bootstrapping technique implemented, respectively. In the first case, uncertainties reflect 1-$\sigma$ errors derived from the covariance matrix when fitting a Gaussian model to measured beam profiles. For bootstrapping, uncertainties reflect the standard deviation of measurements over 10,000 trials. The dashed circle shows the GBT's beam size at 1.4\,GHz, $\theta_{\rm FWHM}=8.2\arcmin$.}
\label{fig:recovered}
\end{figure}

\subsection{Localization Precision}
Three of the four test sources were easily localized to within the GBT's $\theta_{\rm HPBW}$ at 1.4\,GHz (inside the dashed circle in Figure \ref{fig:recovered}) with uncertainties in both $\Delta\alpha_{\rm rec}$ and $\Delta\delta_{\rm rec}\approx1$\,\arcmin. For these three sources (J1400$-$1431, J1930$-$1852, and J1822+0155), recovered positions were mostly consistent with actual values to within 1-$\sigma$ uncertainties and angular separations from timing positions were $\lesssim1\arcmin$. 

Notably, the brightest test source chosen (B1919+21) produced worse results in two separate scans. In the first, we used a faster slew rate (0.666\,\arcmin\,s$^{-1}$), but did not alter the catalog source position; for the second, we used the same slew rate as for the other sources (0.133\,\arcmin\,s$^{-1}$), but changed the center position significantly. In both B1919+21 scans, the uncertainties in measured offsets were much smaller than those for the other sources, which is to be expected given that PSR B1919+21 is much brighter than other test sources included; for comparison, B1919+21 has a flux density measured at 400\,MHz of $S_{400}=57(8)$\,mJy \citep{lyl+95}, while J1400$-$1431 has $S_{350}=4$\,mJy \citep{skm+17}, and other test sources have $S_{350}\lesssim1$\,mJy \citep{rsm+13,srm+15}. However, recovered positions from B1919+21 test scans showed larger separations compared with the pulsar's timing position ($3-4\arcmin$) than separations measured in other scans ($1-2\arcmin$).

Further analysis of B1919+21's test scans revealed no significant discrepancies between the actual antenna pointing positions (diagnostic information recorded as a function of time for all GBT scans) and those calculated using the start position in the header of each pulsar data file. Other systematics (see \S\ref{sec:systematics}) did not help explain the poorly measured position for B1919+21. Fainter sources more similar to new discoveries ($S_{350}\lesssim1$\,mJy) showed very good agreement to withing uncertainties between recovered and timing positions.

In this study, we recovered positions for sources with different flux densities using a OTF mapping technique, fitting a 2D Gaussian to intensities measured every $4.8\arcmin$ at 350\,MHz (30 subintegrations across $2^{\circ}$), and calculated uncertainties in measured beam profile offsets with bootstrapping. For a similar use case, \cite{condon+97}, shows that uncertainties in measured offsets scale like $\sigma_{\Delta\alpha,\Delta\delta}=\sqrt{2/\pi}\,h\,\mu/A$, where $h$ is the spacing between beam profile measurements, $\mu$ is the RMS amplitude uncertainties, and $A$ is the peak amplitude. In practice, we found that this relationship underestimated absolute magnitudes of offset uncertainties by a factor of $3-4$ when compared with those calculated from bootstrapping and simulated data. However, the proportionality can be used to estimate uncertainties resulting from changes in the observing plan (e.g. spacing between subintegrations, slew rate, observing frequency, etc.).

\subsection{Systematic Errors}\label{sec:systematics}
The precision of OTF mapping measurements depends on careful consideration of several systematics that come into play when using the technique.

Persistent radio frequency interference (RFI) may require that large chunks of data be zapped before proceeding with the processing steps outlined in \S\ref{sec:reduction}. We simulated the effects of RFI by ignoring random subsets of the beam profile before conducting 1D/2D fitting to determine position offsets. Even for relatively faint sources, ignoring up to 25\% did not have significant adverse effects on position measurements; therefore, our method can tolerate removal of similar amounts of data due to RFI or potentially, intermittent signals from sources like nulling pulsars and rotating radio transients \citep[RRATs;][]{mll+06}. For bright sources like B1919+21, results are largely unaffected even when up to half the data is discarded. Above these thresholds, measured offsets and uncertainties get progressively larger and eventually, bootstrapping trials fail when less unique beam profile data points become available for fitting.

In addition to nulling, other intrinsic/extrinsic signal to noise variations on short, $\lesssim15$\,min timescales (e.g. diffractive scintillation, slewing across regions with large sky temperature gradients, system temperature variability, etc.) can adversely affect the precision of OTF mapping.

Radio telescopes typically have some amount of intrinsic pointing error, usually corrected by periodically observing a calibrator source in situations when very precise pointing is necessary. Left uncorrected, the GBT's intrinsic pointing error is $5-10\arcsec$,\footnote{This and several figures that follow are taken directly from the GBT observer's guide (\url{https://science.nrao.edu/facilities/gbt/observing/GBTog.pdf}) and private communication with staff on-site.} which is of no concern for routine 350\,MHz observing since $\theta_{\rm HPBW, 350}$ is significantly larger. This level of blind pointing accuracy acceptable for OTF mapping since the measurement precision is larger by approximately an order of magnitude (see Table \ref{tab:offsets}). Time of day and wind speed can also affect tracking accuracy; in the daytime with high-speed ($35-40$\,mph), persistent winds, tracking accuracy may be degraded by $20\arcsec$.

There is a $\approx5$\,s lag between the start of a scan and the moment GUPPI begins writing data to disk, corresponding to a $\approx0.6\arcmin$ discrepancy between the start position of each slew and the {\it actual} start position written to the data header in each case. We accounted for this systematic by calculating recovered positions relative to the position listed in the data header, corresponding to the start of data collection. Another discrepancy was found between the target center position of each map and the actual center position ($\xi$; see Table \ref{tab:offsets}), which varied in magnitude between $0.2-0.7\arcmin$. However, since $\xi$ showed no preferred direction in $\alpha$/$\delta$ (see Figure \ref{fig:recovered}), it is not likely to be indicative of unaccounted-for systematics, but rather the total pointing error due to factors previously mentioned. 

\section{Conclusion}\label{sec:conclusion}
Current single dish, low-frequency radio pulsar surveys sacrifice precise localization of discoveries for efficient sky coverage. Since the telescope's beam size is inversely proportional to the observing frequency, poor localization hampers high-frequency follow-up and degrades the signal-to-noise ratio in future detections. Positions acquired through pulsar timing are extremely precise (often known to a fraction of an arcsecond), but measuring positions this way requires $\simeq1$\,year of dedicated pulsar timing efforts. Therefore, rapid $\approx1\arcmin$ localization is useful for improving the quality (signal-to-noise) of future observations, aids in establishing phase connection, and provides flexibility in follow-up observing frequency, which is useful for evaluating MSP suitability for PTAs. In the current regime, PTA sensitivity is most significantly improved by adding new pulsars to the array \citep{sej+13,tve+16}. The sooner MSPs can be added to PTAs, the more their timing baselines can be extended, maximizing PTA sensitivity to gravitational waves.

Although gridding works well for incremental improvements in pulsar localization, OTF mapping simplifies the procedure when higher precision is required (e.g. improving localization from an $18\arcmin$ error radius to $\approx1-3\arcmin$) for several reasons. First, the observing set-up is nearly identical to that of the discovery scan with more time spent effectively on-source, making follow-up detections more likely (gridding requires assumptions be made about a pulsar's spectral index to ensure detection at higher frequencies). Unlike with gridding, OTF mapping only involves use of a single receiver, eliminating the need for proposing time at multiple frequencies and overhead time for switching between corresponding receivers. Finally, OTF mapping can be used directly on data from drift scan surveys, providing immediate localization in right ascension for free. This is immediately applicable to new discoveries in the 327\,MHz Arecibo drift scan survey \citep{dsm+13} and may also prove useful for current and future drift scan surveys with the Five hundred meter Aperture Spherical Telescope \citep[FAST; e.g.][]{slk+09,zhl+16,lp+16,yln+13}.

This study demonstrates that on-the-fly (OTF) mapping has the capability of localizing pulsars to within $\approx1-3\arcmin$ precision in 30\,mins, with fewer complications and requiring $2-3$ times less telescope time than a traditional gridding method. We also provide code (described in \S\ref{sec:reduction}) to measure position offsets for general purpose use, given standard folded output files from {\tt PRESTO}. The majority of test scans showed reliable position recovery to within measurement uncertainties. For PSR J1400$-$1431, position improvements from OTF mapping facilitated establishing a coherent timing solution, now spanning $>5$\,years \citep{skm+17}. 

\section{Acknowledgements}
We thank David Kaplan, Ryan Lynch, and Scott Ransom for useful comments and suggestions.
We thank West Virginia University for its financial support
of GBT operations, which enabled observations for this project.
The Green Bank Observatory is a facility of the National Science Foundation operated under cooperative agreement by Associated Universities, Inc. Support for JKS and PAG was provided by the NANOGrav NSF Physics Frontiers Center award number 1430284.

\facility{GBT}
\software{astropy \citep{astropy}, scipy \citep{scipy}, {\tt PRESTO} \citep{rem+02}}.

\end{document}